\documentclass{ws-ijmpa}
\usepackage[sort&compress,numbers]{natbib}
\usepackage{graphicx}
\usepackage{stackrel}

\def\ltap{\raisebox{-.55ex}{\rlap{$\sim$}} \raisebox{.4ex}{$<$}}
\def\gtap{\raisebox{-.55ex}{\rlap{$\sim$}} \raisebox{.4ex}{$>$}}
\def\gsim{\mathrel{\gtap}}
\def\lsim{\mathrel{\ltap}}

\begin{document}
\markboth{S.V.~Demidov}{New physics from atmosphere: light sgoldstino case.}

%
\catchline{}{}{}{}{}
%

\title{New physics from atmosphere: light sgoldstino case.}

\author{S.~V.~Demidov}

\address{Institute for Nuclear Research of the Russian Academy of Sciences, \\
  Moscow 117312, Russia, \\
  \vspace{0.1cm}
Moscow Institute of Physics and Technology, Dolgoprudny 141700, Russia, \\
  \vspace{0.1cm}
  Moscow State University, Moscow 119991, Russia \\
    \vspace{0.1cm}
demidov@ms2.inr.ac.ru}

\maketitle

\begin{history}
\received{Day Month Year}
\revised{Day Month Year}
\end{history}

\begin{abstract}
In a supersymmetric model with low scale supersymmetry breaking light
sgoldstinos can appear in decays of mesons abundantly produced in atmospheric 
showers. We obtain bounds on parameter space of such a scenario from the
Super-Kamiokande atmospheric neutrino oscillation data and estimate future
sensitivity of the Hyper-Kamiokande project. We find that the bounds from the
Super-Kamiokande results are weaker than constraints from existing
searches by other experiments while the Hyper-Kamiokande detector can probe a
still allowed part of the model parameter space.

\end{abstract}

\ccode{PACS numbers:}


\section{Introduction}	

Long-awaited new physics may reveal itself at ongoing and near-future particle
physics experiments in the form of new (sub)GeV scale very weakly (feebly)
interacting
particles~\cite{Lanfranchi:2020crw,Agrawal:2021dbo,Antel:2023hkf}. There are
plenty of beyond Standard Model (BSM) scenarios which predict existence of
such type of particles, including models with dark matter, new gauge bosons,
dark photon, axion-like particles, heavy neutral leptons as well as
supersymmetric models. Currently, there is a great interest to (sub)GeV new
physics, partly because no significant evidence for heavy new particles has
been obtained so far at the largest LHC experiments, i.e. ATLAS and CMS.
Weakness of interactions of the hypothetical new light particles to the SM
ones often implies their decay length to be macroscopic. A large experimental
program aimed to probe scenarios with light long-lived particles (LLP) is
underway and continues to develop~\cite{Curtin:2018mvb,Alimena:2019zri}. 

Production mechanisms of new light particles are models dependent and in
many scenarios they are dominantly appear via interactions with mesons which
are abundantly produced in proton collisions. Recently, there has been a lot
of interest in exploring the
possibility~\cite{Kusenko:2004qc,Asaka:2012hc,Masip:2011qb,Arguelles:2019ziu}
of using cosmic rays as a beam and the atmosphere as a dump. In
this case LLPs can be produced in decays of mesons propagating in 
atmosphere, reach a detector and give a signal in it. Prospects of such
searches for several types of BSM models have been
studied~\cite{Coloma:2019htx,Archer-Smith:2020hqq,Plestid:2020kdm,ArguellesDelgado:2021lek,Candia:2021bsl,Arguelles:2022fqq,Cheung:2022umw,Fischer:2023bfn}
including heavy neutral leptons, millicharged particles, axion-like particles,
the Higgs-portal scalar as well as neutralinos in R-parity violating
supersymmetry.   

In the present paper we consider another scenario in which LLPs appear
from a sector responsible for spontaneous supersymmetry breaking. In this type
of supersymmetric models it is assumed that apart from the SM sector the low
energy theory at the electroweak scale contains also light goldstino
supermultiplet. Goldstino, which is Goldstone fermion of spontaneously broken
supersymmetry can be in the simplest case member of a chiral
supermultiplet. The latter contains, apart from goldstino, scalar and
pseudoscalar scalar fields -- sgoldstinos, 
$S$  and $P$, as well as an auxiliary field whose vacuum expectation value
$F$ triggers spontaneous supersymmetry breaking.
The quantity $\sqrt{F}$ 
determines the size of supersymmetry breaking scale. Interactions of
superpartners to the SM fields are fixed by supersymmetry and are generally
related to the SM coupling constants. On the other hand, couplings of the
goldstino sector to visible particles are governed by the supersymmetry
breaking scale and determine the soft supersymmetry breaking parameters.
Interactions of goldstino and sgoldstinos to the SM fields are suppressed by
negative powers of $F$. In particular, sgoldstinos have coupling constants to
the SM sector which scale generically~\cite{Brignole:1996fn,Brignole:2003cm} as
$M_{soft}/F$, where $M_{soft}$ stands for soft supersymmetry breaking
parameters. When supersymmetry breaking scale is much larger than
electroweak scale, i.e. $\sqrt{F}\gg M_{EW}$, goldstino supermultiplet
decouples from usual MSSM fields. Otherwise, if $\sqrt{F}$ is not very far
from TeV scale one should include goldstino sector to the low energy
theory. The latter scenario is called low scale supersymmetry 
breaking and it can be realized in several classes of supersymmetric 
models~\cite{Ellis:1984kd,Ellis:1984xe,Giudice:1998bp,Dubovsky:1999xc}.
In supergravity scenarios goldstino  becomes longitudinal component of
gravitino acquiring the mass $m_{3/2}=\sqrt{8\pi/3}F/M_{Pl}$ via the
super-Higgs mechanism. On the other hand, masses of sgoldstinos depend
on details of the hidden sector and in what follows we consider them to be of
sub-GeV scale. Phenomenology of light sgoldstinos was extensively
studied in
literature~\cite{Gorbunov:2000th,Gorbunov:2000cz,Tchikilev:2003ai,Gorbunov:2005nu,Demidov:2006pt,Demidov:2011rd,Demidov:2022ijc,Sadovsky:2023atd}.  
Here we explore a possibility that sub-GeV sgoldstinos are produced in decays
of kaons in atmospheric showers reach the Super-Kamiokande detector and decay
generating a signal there. We use the strategy advocated in
Refs.~\cite{Arguelles:2019ziu,Coloma:2019htx} and the atmospheric neutrino
oscillation results by Super-Kamiokande~\cite{Super-Kamiokande:2017yvm} to
place bounds on parameter space of 
the model in question as well as to estimate sensitivity reach of the 
Hyper-Kamiokande project~\cite{Hyper-Kamiokande:2018ofw} to the same type of
signal.  

The rest of the paper is organized as follows. Section~2 contains concise
description of the model with light sgoldstinos and discussion of the relevant 
sgoldstino production and decay modes. In Section~3 we discuss methods used to
estimate the flux of sgoldstinos produced in atmosphere and describe the
obtained results. Section~5 contains our
conclusions.   

\section{SUSY with light sgoldstino}
We assume sgoldstino to be of GeV scale and consider other superpartners as
decoupled. Below we present the relevant part of the interaction lagrangian of
sgoldstinos to the SM fields (see~\cite{Gorbunov:2001pd} for detailed
derivation). The interactions of sgoldstino with the gauge fields read 
\begin{equation}
  \label{eq:2.1}
{\cal L}_{gauge} = -\sum_{\alpha}\frac{M_{\alpha}}{2\sqrt{2}F}
{S}F_{a\;\mu\nu}^{\alpha} F_{a}^{\alpha\;\mu\nu}
-\sum_{\alpha}\frac{{M_{\alpha}}}{2\sqrt{2}F}
{ P}F_{a\;\mu\nu}^{\alpha} \tilde{F}_{a}^{\alpha\,\mu\nu}\,,
\end{equation}
where $\tilde{F}^{\alpha\,\mu\nu} =
\frac{1}{2}\epsilon^{\mu\nu\lambda\rho}F^{\alpha}_{\lambda\rho}$ and
$M_\alpha, \alpha=1,2,3$ are gaugino masses, while the interactions with
quarks and charged leptons are given by
\begin{equation}
  \label{eq:2.2}
{\cal L}_{fermions} =
-\frac{{A^q_{ab}v}}{\sqrt{2}F}S\bar{q}_{a}q_{b}  - i\frac{{A^q_{ab}v}}{\sqrt{2}F}P\bar{q}_{a}\gamma^5q_{b} 
-\frac{{A^l_{ab}v}}{\sqrt{2}F}S\bar{l}_{a}l_{b}  - i\frac{{A^l_{ab}v}}{\sqrt{2}F}P\bar{l}_{a}\gamma^5l_{b}\,. 
\end{equation}
Here $A_{ab}^{q}$ and $A_{ab}^{l}$ are parameters, directly related to soft
trilinear coupling constants in MSSM, $v$ is the Higgs boson vacuum
expectation value. In what follows we assume the following hierarchy
$A_{ab}^{q,l}v=m_a^{q,l}A_0\delta_{ab}$, where $m_a^{q,l}$ is the quark or
lepton mass and $A_0$ is a common mass scale for trilinear coupling
constants. This assumption leaves us with flavour violating sources pertinent
to the Standard Model. Apart from the interactions described by~\eqref{eq:2.1},
\eqref{eq:2.2} the scalar sgoldstino can mix with the Higgs
boson~\cite{Astapov:2014mea} with the 
mixing angle given by 
\begin{equation}
  \label{eq:2.3}
\theta_{Sh} \approx -\frac{4{\mu}^3v\sin{2\beta} +
  v^3(g_1^2{M_1}+g_2^2{M_2})\cos{2\beta}}{2F m_h^2}\,,
\end{equation}
where $\mu$ and $\tan{\beta}$ are standard parameters of MSSM.
If the mixing dominates production of light scalar sgoldstino the
phenomenology of this scenario is equivalent to that of the Higgs portal
model. The 
corresponding signal from atmosphere was discussed
recently~\cite{Archer-Smith:2020hqq}. In what follows we 
consider parameter space of the model where contribution of the
sgoldstino-Higgs mixing is small. 

The model describing interactions of sgoldstino with the SM
fields~\eqref{eq:2.1},\eqref{eq:2.2} should be considered as an effective
field theory valid at energies $E\lsim \sqrt{F}$. Higher order interactions
are suppressed by higher powers of $F$. Weak coupling regime of this theory
implies hierarchy $M_{soft}\lsim \sqrt{F}$ (unitarity bound) . Direct searches
for the 
superpartners at LHC experiments push bounds on their masses to around and
beyond TeV scale~\cite{ATLAS:2024exu} and so does the bound on the
supersymmetry 
breaking scale $\sqrt{F}$.

The parameter space of the model considered in this study corresponds to
sgoldstino decaying mainly into photon pair, $e^+e^-$ and $\mu^+\mu^-$
pair. Previous 
studies~\cite{Gorbunov:2000th,Demidov:2022ijc} show that this is the  case
for scalar (pseudoscalar) 
sgoldstino of mass $m_S < 2m_\pi$ ($m_P < 3m_\pi$) as at larger masses hadron
decay channels starts to dominate. Partial widths of sgoldstino decays
into pair of charged leptons read  
\begin{equation}
  \label{eq:2.4}
  \Gamma(S(P)\to l\bar{l}) = \frac{{A_0}^2m_{l}^2m_{S(P)}}{16\pi F^2}\sqrt{1-\frac{4m_l^2}{m^2_{S(P)}}}\,,
\end{equation}
where again the hierarchy $A_{aa}^lv=m_a^lA_0$ is assumed. 
The partial width of scalar sgoldstino decay into pair of photons is given by 
\begin{equation}
  \label{eq:2.5}
  \Gamma(S\to\gamma\gamma)= \frac{{M_{\gamma\gamma}}^2 m_{S}^3}{32\pi F^2}\,,
\end{equation}
where $M_{\gamma\gamma}\equiv M_1\cos^2{\theta_W}+M_2\sin^2{\theta_W}$ with
$\theta_W$ being Weinberg angle. For pseudoscalar sgoldstino the amplitude of
$P\to\gamma\gamma$ decay receives additional contributions from the mixing
with light pseudoscalar mesons~\cite{Gorbunov:2000th} and using Chiral
Perturbation Theory (ChPT) one can obtain 
\begin{equation}
  \label{eq:2.6}
  \Gamma(P\to\gamma\gamma)=\frac{m_{S}^3}{32\pi F^2}\left(M_{\gamma\gamma} -
    \frac{4\alpha M_3}{3\alpha_s(M_3)}\,
      \frac{5m_K^2-3m_P^2-2m_\pi^2}{4m_K^2-3m_P^2-m_\pi^2} - \frac{\alpha
      A_0}{\pi}\right)^2\,,
\end{equation}
where $\alpha_s(M_3)$ is strong coupling constant evaluated at scale $M_3$
(see a discussion in Ref.\cite{Gorbunov:2000th}) and
we neglect small scale dependence of fine structure constant $\alpha$. 
For simplicity, in the present study we disregard
isospin violating contributions as well as $\eta$--$\eta^\prime$ mixing
effects. 

As for light sgoldstino production mechanisms in atmospheric showers we
take into account sgoldstino production in kaon decays. The following decay
channels are relevant: $K^\pm\to\pi^\pm S$, $K_L\to
\pi^0 S$ for scalar sgoldstino and $K^\pm\to \pi^\pm P$, $K_S\to \pi^0 P$,
$K^\pm\to P \mu\nu_\mu$ and $K^\pm\to P e\nu_e$ for pseudoscalar
sgoldstino. Decays $K^\pm\to S\mu\nu_\mu$ and $K^\pm\to S e\nu_e$ are found to
be always subdominant as compared to $K\to \pi S$ for $m_S<2m_\pi$ and we
neglect them  in what follows. 

The amplitudes of $K^\pm\to\pi^\pm S$ and $K_L\to \pi^0 S$ decays can be
obtained from the ChPT results for $K\to \pi H$ decays~\cite{Leutwyler:1989xj}
with a suitable redefinition of coupling constants and read
\begin{align}
  \nonumber
{\cal A}(K^\pm&\to\pi^\pm S) = -{\cal A}(K_L\to\pi^0 S) = \\ 
= &
    \frac{m_K^2}{2}\left[\left(\gamma_1\left(1-\frac{m_S^2-m_\pi^2}{m_K^2}\right)
    -  \gamma_2\right)\left(\frac{A_0}{\sqrt{2}F} + \frac{2\sqrt{2}\pi 
    M_3}{9\alpha_s(M_3)}\right) + \zeta\frac{A_0}{\sqrt{2}F}\right]\,,
    \label{eq:2.7}
\end{align}
where $\zeta = \frac{3\sqrt{2}G_F}{16\pi^2}\sum_{i=c,t}
V_{id}^*V_{is}m_i^2\simeq 3.3\cdot 10^{-6}$,
$\gamma_1\simeq 3.1\cdot10^{-7}$ and $|\gamma_2/\gamma_1|\ll 1$.
For pseudoscalar sgoldstino the amplitudes of $K\to P\pi $ decays are given by 
\begin{align}
  \nonumber
  {\cal A}(K^\pm&\to\pi^\pm P) = -{\cal A}(K_S\to\pi^0 P) = 
  -2G_8 F_\pi^2\frac{m_K^2-m_\pi^2}{4m_K^2-3m_P^2-m_\pi^2} \times \\
 & \times \left(\frac{A_0}{\sqrt{2}F}(2m_K^2 + m_\pi^2 - 3m_P^2) +
   \frac{4\sqrt{2}\pi M_3}{\alpha_S(M_3) F}(m_K^2 - m_P^2)\right) \,,
   \label{eq:2.8}
\end{align}
where $F_\pi = 0.093$~GeV and $G_8 =
-\frac{G_F}{\sqrt{2}}V^*_{ud}V_{us}g_8$ with $g_8\simeq
5.0$~\cite{Cirigliano:2011ny}. For the amplitudes of semileptonic kaon decays
involving pseudoscalar sgoldstino we obtain
\begin{equation}
  \label{eq:2.9}
  {\cal A}(K^\pm \to P l\nu_l) = -{\cal A}(K^\pm \to \pi^0 l\nu_l)
   \frac{3\sqrt{2}F_\pi(m_K^2 - m_\pi^2)}{4m_K - 3m_P^2 -
    m_\pi^2}\left(\frac{A_0}{F} + \frac{4\pi M_3}{3\alpha_s(M_3) F}\right)\,,
\end{equation}
where ${\cal A}(K^\pm \to \pi^0 l\nu_l)$ is the amplitude of $K^\pm \to \pi^0
l\nu_l$. Some details on derivation of the above expressions for the case of 
pseudoscalar sgoldstino are presented in Appendix~A.

Given the sgoldstino interaction lagrangian~\eqref{eq:2.1},~\eqref{eq:2.2}
and the assumptions about parameters of the model discussed above all the
relevant sgoldstino production decay widths are governed by 
sgoldstino masses as well as two parameter ratios, $\frac{M_3}{F}$ and
$\frac{A_0}{F}$ which determines interaction of sgoldstinos to gluons and
quarks, respectively.  

Concluding this Section let us note that there is a possibility of decays of
light mesons into final states containing pair of
sgoldstinos~\cite{Gorbunov:2000th,Demidov:2011rd} but corresponding 
amplitudes are expected to be suppressed by additional power of $M^2_{soft}/F$
ratio and we neglect them in what follows.

\section{Signal from light sgoldstinos: bounds from Super-Kamiokande and
  sensitivity of Hyper-Kamiokande}
Light sgoldstinos can be copiously produced from kaon decays in atmospheric
showers and then decay into photons producing a signal in a neutrino 
detector. In simulation of production of new light particles in meson decays
in atmosphere we follow procedure proposed and used in several recent
studies~\cite{Arguelles:2019ziu,Coloma:2019htx,Archer-Smith:2020hqq,Candia:2021bsl,Arguelles:2022fqq,Cheung:2022umw}.
The flux $\Phi$ of sgoldstino produced in meson decays in atmosphere can be 
calculated as~\cite{Gondolo:1995fq}
\begin{equation}
  \label{eq:3.1}
  \frac{d\Phi}{dEd\Omega dX}(E,X) = \sum_{M}\int dE_M\;
  \frac{1}{\rho(X)\lambda_M(E_M)}\frac{d\Phi_M}{dE_Md\Omega}(E_M,X)\frac{dN}{dE}(E_M,E)\,.
\end{equation}
Here $\frac{d\Phi_M}{dE_Md\Omega}$ and $\rho$ are the differential flux of
meson $M$ and the density of atmosphere at a column depth $X$, respectively,
$\lambda_M\equiv\gamma_M\beta_Mc\tau_M$ is the decay length of meson $M$ with the
boost factor $\gamma_M\beta_M$ and proper lifetime $\tau_M$. Finally,
$\frac{dN}{dE}(E_M,E)$ is the energy spectrum of sgoldstino produced in the
decays of meson $M$ having energy $E_M$. The sum in~\eqref{eq:3.1} goes over
all mesons contributing sgoldstino production.

\begin{figure}[!htb]
  \centerline{\includegraphics[width=7.5cm]{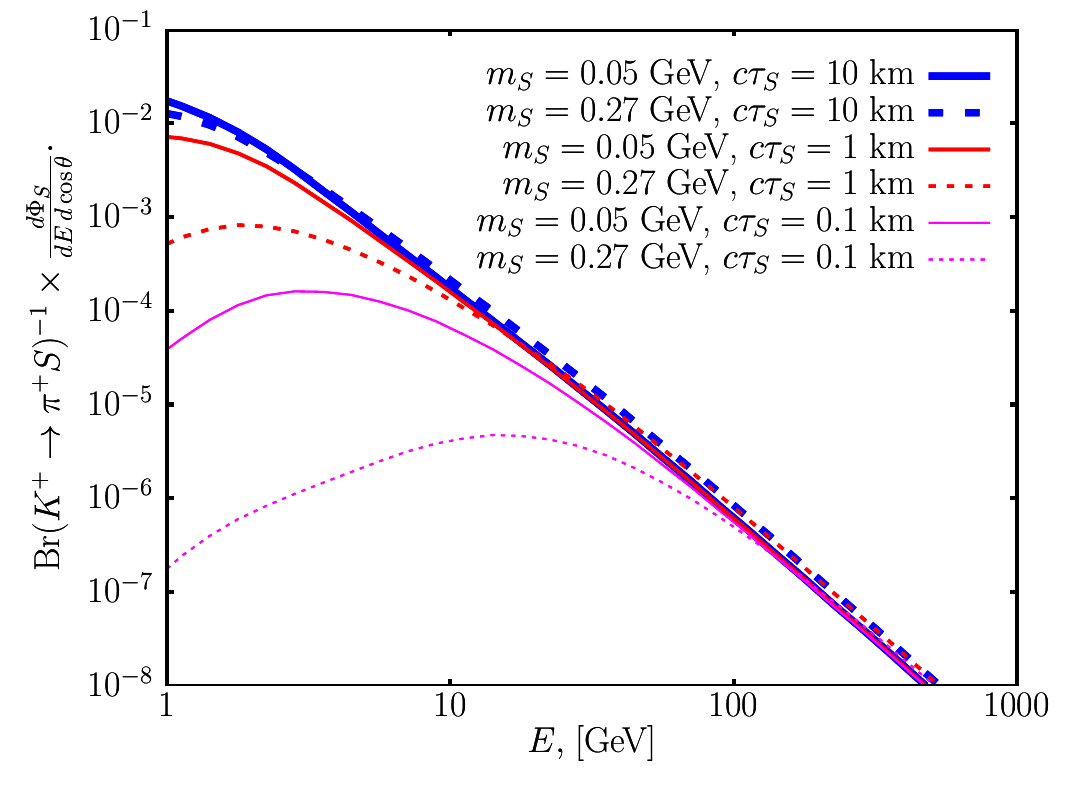}}
\caption{Normalized fluxes of scalar sgoldstino at the detector level for $m_S
  = 0.05$~GeV and 0.27~GeV and several values of $c\tau$. \label{f1}}
\end{figure}

To simulate production rate of mesons in atmosphere we use Matrix Cascade
Equations (MCEq) program~\cite{Fedynitch:2012fs,Fedynitch:2015zma}, which
allows to solve numerically the system of cascade
equations~\cite{Gondolo:1995fq} which governs the fluxes of particles produced in
atmospheric showers. We use the Hillas-Gaisser cosmic ray
model~\cite{Gaisser:2011klf} and NLRMSISE-00 atmospheric
model~\cite{nrlmsise-00}. As for hadronic interaction models, most 
of the results below are obtained with SYBILL-2.3c~\cite{Fedynitch:2018cbl} and
to estimate an uncertainty related to hadronic models we make a comparison
with those obtained using QGSJET-II-04~\cite{Ostapchenko:2010vb} and
DPMJET-III~\cite{Roesler:2000he}. We have implemented the energy distribution
$\frac{dN}{dE}$ of sgoldstino produced in decays of boosted kaons in MCEq as follows
\begin{equation}
  \frac{dN}{dE}(E_M,E) =
  \int_{E_{0}^{min}}^{E_{0}^{max}}\,dE_0\frac{1}{2\gamma_M\beta_M\sqrt{E_0^2 - m_{S(P)}^2}}\frac{dN_0}{dE_0}(m_M,E_0)\,,
\end{equation}
where $\frac{dN_0}{dE_0}$ is the sgoldstino energy distribution in the kaon rest frame
calculated using the decay amplitudes discussed in the previous Section and
\begin{equation}
  E_{0}^{min} = \gamma_M\left(E - \beta_M\sqrt{E^2-m^2}\right)\,,\;
  E_{0}^{max} = \gamma_M\left(E + \beta_M\sqrt{E^2-m^2}\right)\,,
\end{equation}
with $m$ being $m_S$ or $m_P$.
The sgoldstino flux at the detector is given by
\begin{equation}
  \label{eq:3.4}
\frac{d\Phi_S}{dEd\cos{\theta}} = 2\pi \int dX\; {\rm
  e}^{-l/\lambda_{dec}}\;\frac{d\Phi}{dEd\Omega dX}\,,
\end{equation}
where $l$ is the distance sgoldstino travels from the production point to the
detector and $\lambda_{dec} = \gamma\beta c\tau$ is sgoldstino decay length
in the lab frame. We are interested in sgoldstino decay signal in the
Super-Kamiokande detector placed under 1000~m of
rock~\cite{Super-Kamiokande:2002weg} and this underburden is taken into account
in Eq.\eqref{eq:3.4}. In Fig.~\ref{f1} we plot the normalized energy spectra
of scalar sgoldstino  
with mass $m_S=0.05$~GeV and 0.27~GeV at Super-Kamiokande for several values
of $c\tau$. We see that for shorter sgoldstino lifetimes the maximum of the
spectra shifts to higher energies as only sufficiently boosted unstable
particles may reach the detector.  

Sgoldstino can decay inside the instrumented volume of the Super-Kamiokande
detector into a pair of photons each resulting in a showering or $e$-like
event. Similar events from $\pi^0$ decays with two reconstructed Cerenkov
rings are used for calibrations of the 
detector~\cite{Super-Kamiokande:2002weg}. However, most of such events have
momenta less than 1~GeV as for high energy $\pi^0$ the rings
overlap~\cite{Super-Kamiokande:2002weg}. Therefore, to set a conservative
bounds on the model parameter space we consider $S(P)\to\gamma\gamma$ decay as
resulting to a single $e$-like ring and use the Super-Kamiokande
results~\cite{Super-Kamiokande:2017yvm} on the atmospheric neutrino
oscillations.  Namely, we use the data and expected background for $e$-like
fully contained events, both ``Multi-GeV'' and ``Multi-Ring'', binned in
$\cos{\theta}$. 

To calculate number of signal events in Super-Kamiokande we use the effective
area
\begin{equation}
S_{eff}^{SK}(\theta, E) = \int dS_\perp \;\left(1 - {\rm
    e}^{-\frac{\Delta_{det}}{\lambda_{dec}}}\right)\,,
\end{equation}
which was analytically calculated~\cite{Arguelles:2019ziu} for cylinder
geometry of the detector with height of 40~m and radius 20~m.  
The expected number of signal events in each of 10 bins in $\cos{\theta}$ is
calculated as 
\begin{equation}
  \label{eq:3.6}
  S_i = \epsilon T \int d\cos{\theta}\, dE\, S_{eff}^{SK}(\theta, E)\,
  \frac{d\Phi}{dE d\cos{\theta}}\,,
\end{equation}
where $T=5326$~days and we use $\epsilon=0.75$ for detection
efficiency~\cite{Super-Kamiokande:2017yvm}. In Eq.~\eqref{eq:3.6} we integrate
over the energy range 1.33--90~GeV which correspond to that of atmospheric
neutrino for selected sets of events.

To obtain the bounds on the parameters of the model with light sgoldstinos
from the Super-Kamiokande data we use the following procedure. Firstly, we
consider cases scalar and pseudoscalar sgoldstino separately. After fixing the
value of sgoldstino mass we are left with the following free relevant
parameters, namely, $c\tau$ determined by $\frac{M_{\gamma\gamma}}{F}$ and the
couplings which govern sgoldstino production $\frac{A_0}{F}$ and
$\frac{M_3}{F}$. Assuming that one of these couplings gives a dominant
contribution to sgoldstino production rate we use 
\begin{equation}
  \chi^2 = 2\sum_i\left(S_i+B_i-N_i\left(1-\log{\frac{N_i}{S_i+B_i}}\right)\right)\,,
\end{equation}
where $S_i, B_i$ and $N_i$ are number of signal, background and data events in
$i$th bin, respectively, to determine 90\%~CL bound on the left two parameter
space. 

To estimate sensitivity reach of the Hyper-Kamiokande
project~\cite{Hyper-Kamiokande:2018ofw} we calculate the signal from
sgoldstino decay for cylindrical geometry of the detector with the hight of
71~m and diameter of 68~m and exposure corresponding to 20~years of
operation~\cite{Hyper-Kamiokande:2020aij}. The number of background events
has been estimated conservatively by an appropriate scaling from the
Super-Kamiokande case assuming 8.4~time larger fiducial
volume~\cite{Hyper-Kamiokande:2022smq}.  

\begin{figure}[!htb]
  \centerline{\includegraphics[width=6.5cm]{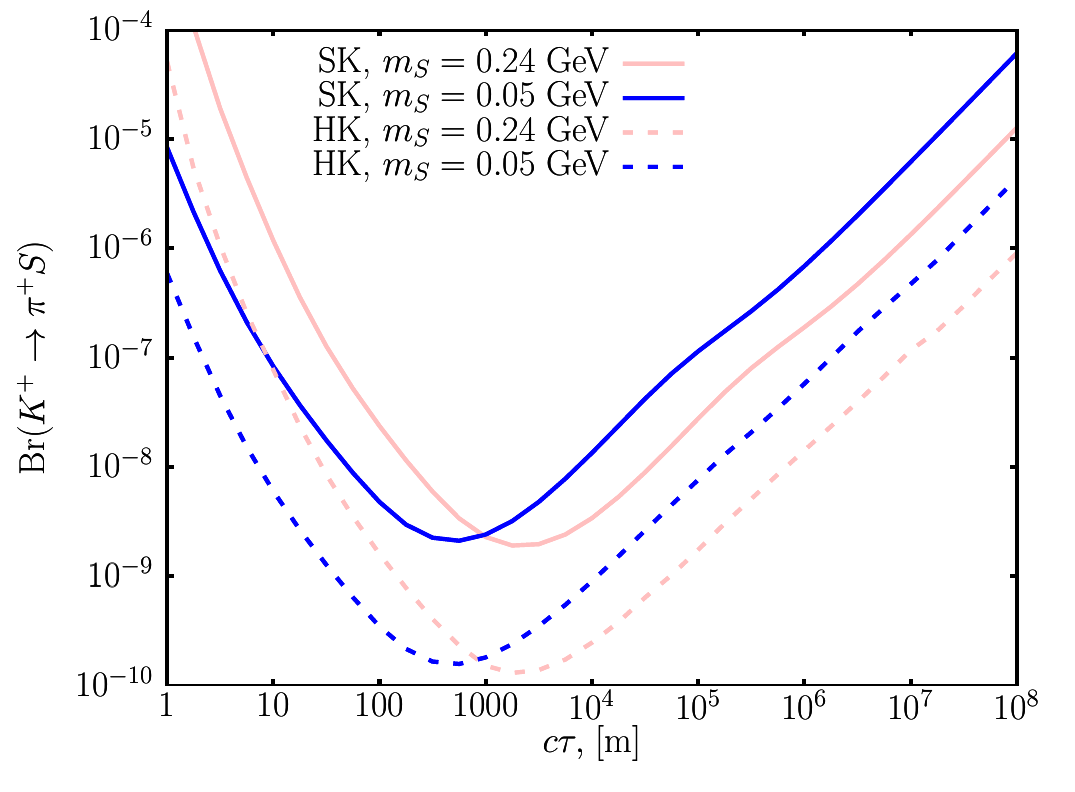}
  \includegraphics[width=6.5cm]{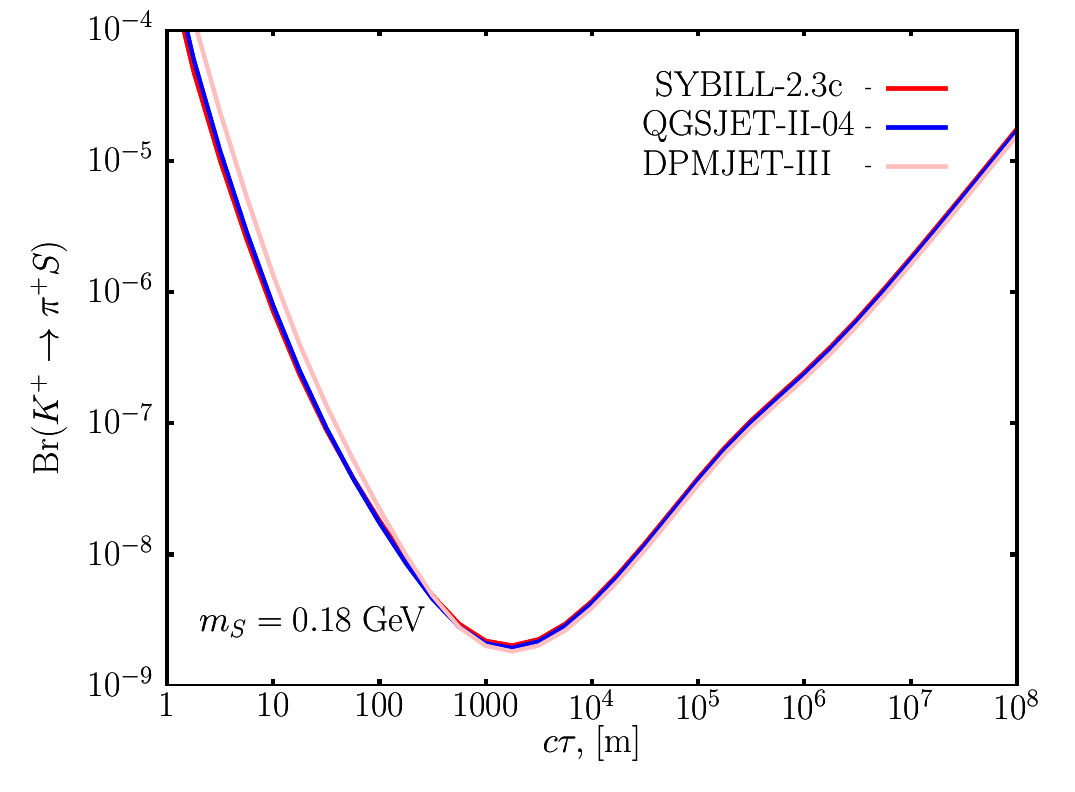}}
\caption{{\it Left panel:} 90\% CL upper limits from the Super-Kamiokande data
  and 
  sensitivity reach of Hyper-Kamiokande to ${\rm Br}(K^+\to \pi^+ S)$ for scalar
  sgoldstino. {\it Right panel:} Comparison of the upper bounds on ${\rm
    Br}(K^+\to \pi^+ S)$ for scalar sgoldstino from the Super-Kamiokande data 
  for different hadronic models. \label{f2}} 
\end{figure}

In Fig.~\ref{f2} (left panel) 
we show the limits from the Super-Kamiokande
data in $(c\tau, {\rm Br}(K^+\to \pi^+ S))$ plane for fixed values of the
scalar sgoldstino masses. We see that the strongest constraints are obtained
for models with $c\tau\sim 10^2-10^4$~m. On the right panel of this Figure we
present the comparison of the upper bounds on ${\rm  Br}(K^+\to \pi^+ S)$ for
scalar sgoldstino obtained with different hadronic models. We obtain that the
uncertainty related to the hadronic models is less than 10~\% for $c\tau\gsim
100$~m. Bounds and sensitivity for smaller values of $c\tau$ apart from larger
value of such uncertainty require precise knowledge of mountain topography
around the detector.

\begin{figure}[!t]
  \centerline{\includegraphics[width=6.5cm]{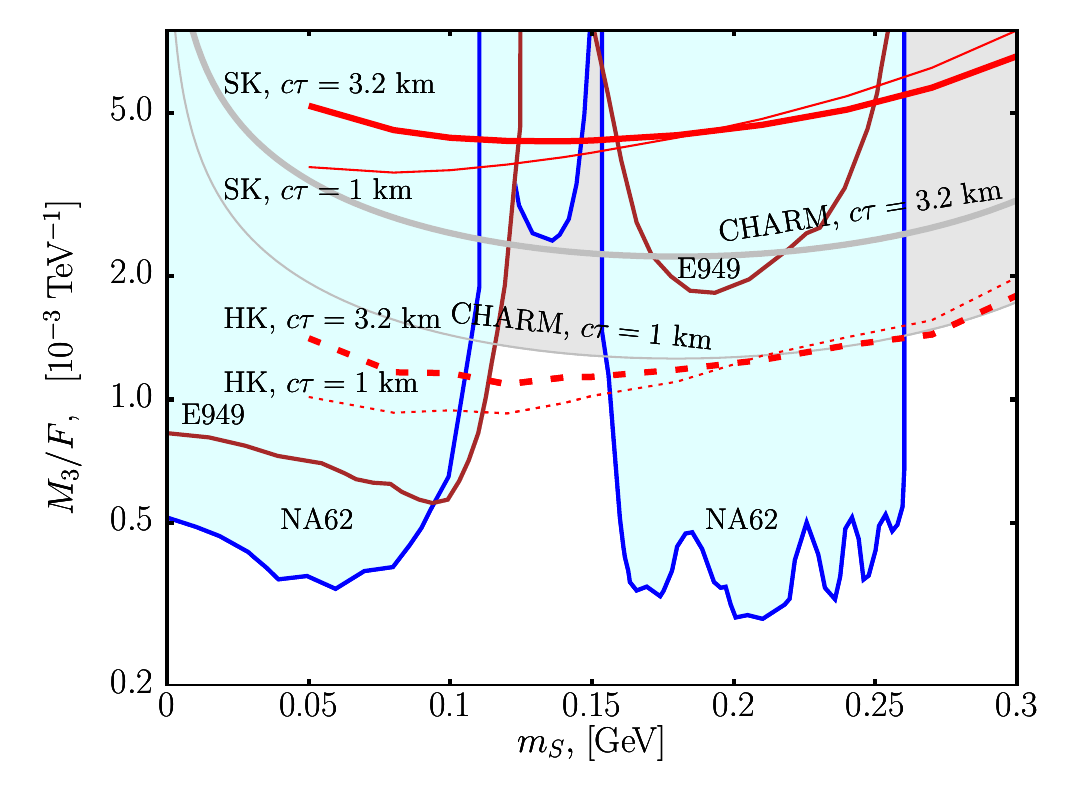}
  \includegraphics[width=6.5cm]{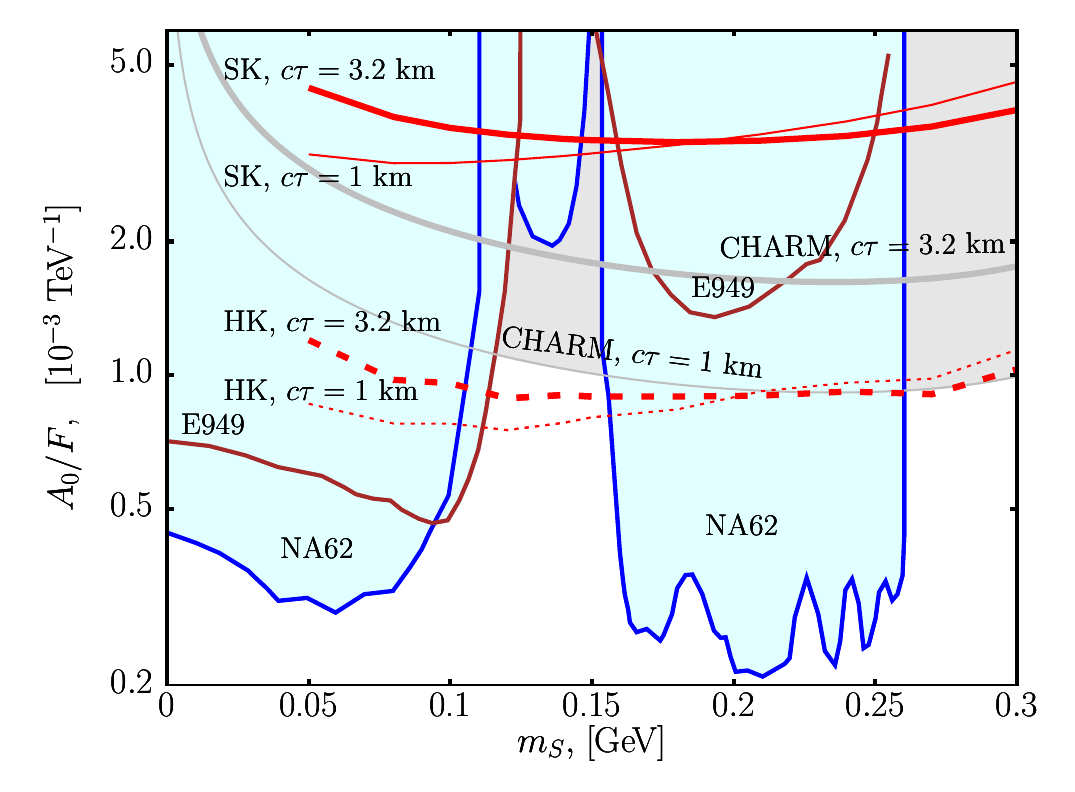}}
\caption{90\% bounds from the Super-Kamiokande results and the Hyper-Kamiokande
  sensitivity reach to parameters of model with light scalar 
  sgoldstino in comparison with the other relevant existing experimental
  constraints (NA62, E494, CHARM). Left (right) panel corresponds to
a scenario in which sgoldstino production in kaon decays is dominated by
coupling to gluons, i.e. $\frac{M_3}{F}$ (quarks, i.e. $\frac{A_0}{F}$). \label{f3}} 
\end{figure}

In Fig.~\ref{f3}
we show the bounds on the model parameter space for the case of light scalar
sgoldstino assuming ${\rm Br}(S\to\gamma\gamma)\approx 1$ from the
Super-Kamiokande data as well as expected sensitivity reach of
Hyper-Kamiokande in comparison with several existing
constraints, namely from NA62~\cite{NA62:2020xlg,NA62:2020pwi},
E949~\cite{BNL-E949:2009dza} and CHARM~\cite{CHARM:1985anb} experiments. To
obtain the bounds from CHARM experiment we follow approach described in
Ref.~\cite{Egana-Ugrinovic:2019wzj}. We see that the parameter space
constrained by the Super-Kamiokande results is already excluded by other
searches. At the same time Hyper-Kamiokande is potentially sensitive to a
still allowed parameter space with scalar sgoldstino masses in the ranges 
0.11--0.15~GeV and 0.26--0.28~GeV. 

\begin{figure}[!htb]
  \centerline{\includegraphics[width=6.5cm]{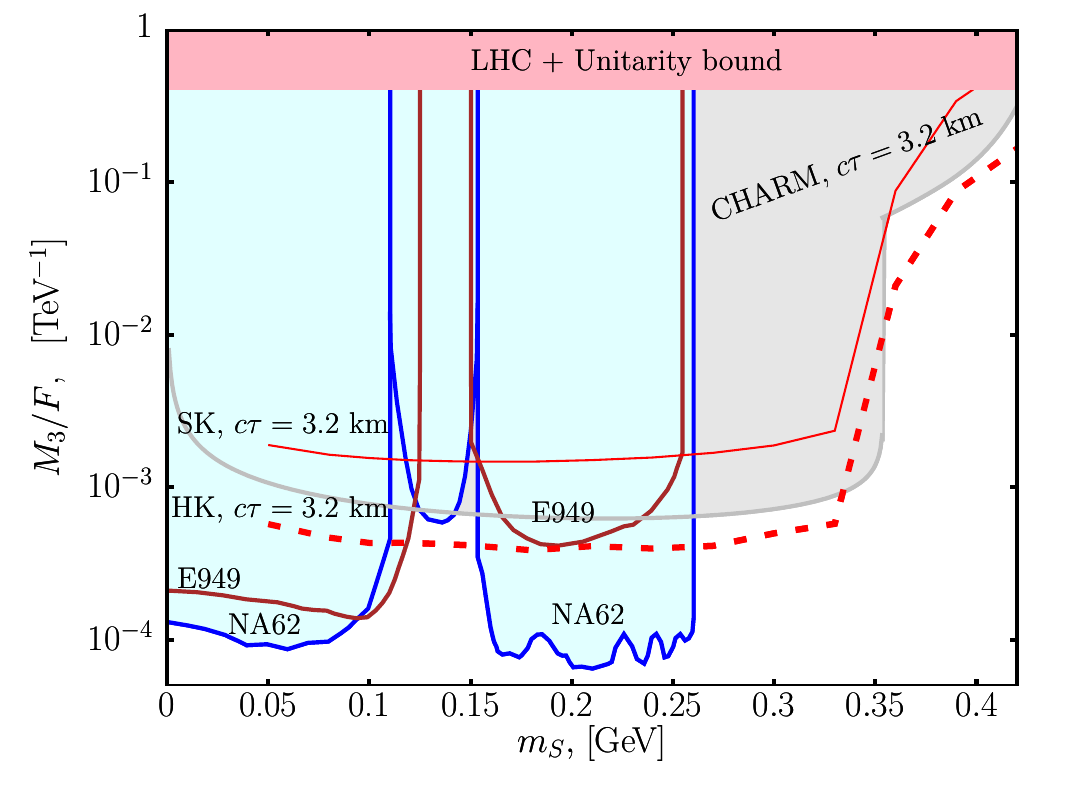}
  \includegraphics[width=6.5cm]{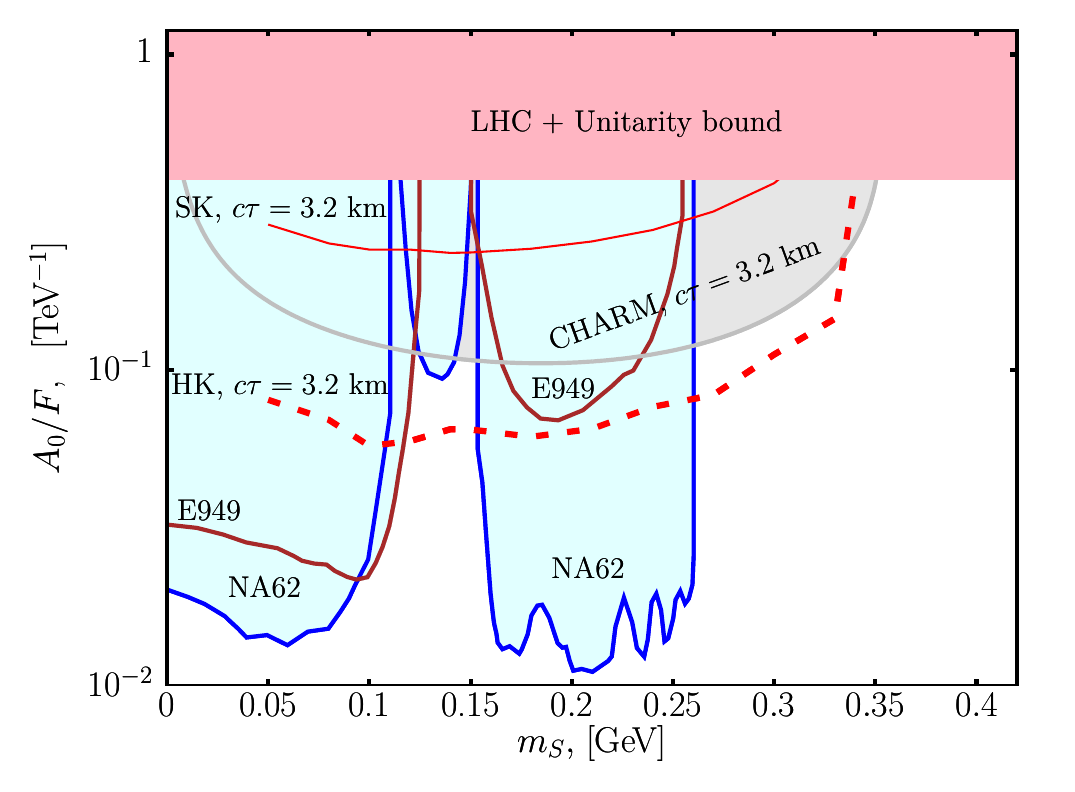}}
\caption{The same as in Fig.~\ref{f3} but for case of pseudoscalar
  sgoldstino. See also main text.  \label{f4}}
\end{figure}

In Fig.~\ref{f4} we show 
the Super-Kamiokande bounds and the Hyper-Kamiokande sensitivity to the model
parameter space for the case of pseudoscalar sgoldstino assuming ${\rm
  Br}(P\to\gamma\gamma)\approx 1$. Apart from existing experimental
constraints from NA62, E494 and CHARM we put in this Figure the unitarity
bound which corresponds to the condition $\sqrt{F}<M_{soft}$ for which the
model is no longer in perturbative regime, taking for $M_{soft}$ a
conservative bound $M_{3}\gsim 2.5$~TeV from the LHC
data~\cite{ATLAS:2024exu}.
Somewhat weaker bounds/sensitivity for the scenario in which sgoldstino
production is dominated by coupling $\frac{A_0}{F}$ on the right panel, as
compared to that of saturated by $\frac{M_3}{F}$ on the left panel, is related 
to a numerical enhancement by factor of order $\frac{\pi}{\alpha_s(M_3)}$, see 
Eqs.~\eqref{eq:2.8}--\eqref{eq:2.9}. The same enhancement for the case of scalar sgoldstino
production is partially compensated by a balance between numerical constants
$\gamma_1$ and $\zeta$ in Eq.~\eqref{eq:2.7}.
As we discussed in Section~2 the pseudoscalar
sgoldstino can decay dominantly into pair of photons for $m_P< 3m_\pi$. For
$m_P\lsim 0.33$~GeV it is produced dominantly via $K^\pm\to \pi^\pm P$ decay
mode while at larger masses the sgoldstino production channels are $K^\pm\to P
l \nu_l$ where $l=e$ and/or $\mu$. Branching fractions of the latter
decays are smaller because of the phase space suppression. This considerably
weakens the constraints from the Super-Kamiokande results and future
sensitivity of Hyper-Kamiokande. 

Let us briefly discuss the parameter space which could be potentially
probed in this type of searches. They are most sensitive to models with
sgoldstino having $c\tau\sim 10^2-10^4$~m. Assuming dominant decay into pair
of photons this corresponds to $\frac{M_{\gamma\gamma}}{F}\sim
10^{-6}-10^{-3}$~TeV$^{-1}$. In the considered scenario the production of
sgoldstino in kaon decays are governed by different couplings,
i.e. $\frac{M_3}{F}$ and $\frac{A_0}{F}$. Current experimental bounds
and the Hyper-Kamiokande sensitivity reach values for these constants up to
$10^{-4}-10^{-3}$~TeV$^{-1}$. The unitarity bound then tells us that searches
for light sgoldstino can probe supersymmetry breaking scale up to
$10^{3}-10^{4}$~TeV. 

\section{Conclusions}
Atmospheric beam dump is an interesting avenue for searches of light
long-lived particles. In this study we obtain bounds on parameter space of the
supersymmetric model with low scale supersymmetry breaking and light
sgoldstinos having their masses at sub-GeV scale from the Super-Kamiokande
atmospheric neutrino oscillation results. Long-lived scalar and pseudoscalar
sgoldstino are assumed to be produced in kaon decays in atmosphere and decay
into pair of photons in the detector. We also estimate future sensitivity of
the Hyper-Kamiokande project to this type of searches and compare the results
to the
existing experimental bounds. We obtain that although the bounds from the
Super-Kamiokande results do not give us new constraints on the model, searches
with the Hyper-Kamiokande detector can probe parts of the model parameter space
which are still avaliable. It is very interesting that searches for new light  
particles produced in atmosphere discussed in this study are complimentary to
future neutrino collider experiments such as MicroBooNE, JSNS$^2$, SBND and
DUNE, see
e.g.~\cite{Coloma:2022hlv,MicroBooNE:2025gpp,Ema:2023tjg,Coloma:2023oxx,Berger:2024xqk}.     

\section*{Acknowledgments}
I am grateful to Dmitry Gorbunov for discussions.
This work is supported in the framework of the State project ``Science'' by
the Ministry of Science and Higher Education of the Russian Federation under
the contract 075-15-2024-541.

\appendix
\section{Sgoldstino production and decays}
In the Appendix we present details on derivation of expressions for the
amplitudes or/and decay widths of 
the decay channels involving sgoldstino which are used in the main part of the
paper. We use the Chiral Perturbation Theory (ChPT) which is most relevant for
kaon decays and sgoldstino with the masses smaller than 0.4~GeV. We closely
follow a 
consistent approach~\cite{Georgi:1986df,Bauer:2020jbp,Bauer:2021wjo} developed  
originally for description of axion-like particle phenomenology.

To determine interactions of pseudoscalar sgoldstino with light mesons one
makes a chiral rotation of the light ($u$, $d$, $s$) quark fields
\begin{equation}
  \label{eq:a:1}
q(x) \to {\rm exp}\left(i\alpha_q(x)\gamma^5\right)\,q(x)\,,
\end{equation}
where $\alpha_q(x) = \frac{\sqrt{2}\pi M_3}{\alpha_s(M_3) F}\kappa_qP(x)$ and
$\kappa_u, \kappa_d, \kappa_s$ are arbitrary constants. This rotation modifies
coupling constants of pseudoscalar sgoldstino not only to quarks but also to
gauge bosons, i.e. gluons and photons, due to chiral anomaly. Namely, the
transformation~\eqref{eq:a:1} results in the following substitution
\begin{gather}
  \label{eq:a:2}
  \frac{M_3}{2\sqrt{2}F} \to
  \frac{M_3}{2\sqrt{2}F}(1-\kappa_u-\kappa_d-\kappa_s)\,, \\
  \label{eq:a:3}
  \frac{M_{\gamma\gamma}}{2\sqrt{2}F} \to
  \frac{M_{\gamma\gamma}}{2\sqrt{2}F}\left(1 -
    2N_c\frac{\alpha}{\alpha_s}\left(\kappa_uQ_u^2 + \kappa_dQ_d^2 + \kappa_sQ_s^2\right)\right)
\end{gather}
in the interaction lagrangian~\eqref{eq:2.1} where $N_c=3$ is the number of
colors 
and $Q_a$ is the electric charge of quark $q_a$. Taking $\kappa_q$ such as
$\kappa_u + \kappa_d + \kappa_s = 1$ eliminates interaction of $P$ with gluons
and makes application of the ChPT straightforward. In our study for simplicity
we neglect $\eta$--$\eta^\prime$ mixing and isospin violating effects. 

The corresponding interaction lagrangian at ${\cal O}\left(p^2\right)$ reads
\begin{equation}
  \label{eq:a:4}
{\cal L}^P = \frac{F_\pi^2}{4}{\rm Tr}\left(\partial
  \Sigma^\dagger\partial\Sigma + \chi\Sigma^\dagger +
  \chi^\dagger\Sigma\right) - G_8F_\pi^4{\rm
  Tr}\left(\Sigma\partial\Sigma^\dagger\Sigma
  \partial\Sigma^\dagger\lambda\right)\,, 
\end{equation}
where $\Sigma = {\rm exp}(-i\hat{\kappa}P)\,U\,{\rm exp}(-i\hat{\kappa}P)$ with
$\hat{\kappa}={\rm diag}(\kappa_u,\kappa_d,\kappa_s)$. We also have
\begin{equation}
  U = {\rm exp}\left(\frac{i\sqrt{2}}{F_\pi}\Phi\right)\,,\;\;\;
  \Phi = \left(\begin{array}{ccc}
                 \frac{\pi^0}{\sqrt{2}}+\frac{\eta}{\sqrt{6}} & \pi^+ & K^+\\
                 \pi^- & -\frac{\pi^0}{\sqrt{2}}+\frac{\eta}{\sqrt{6}} & K^0\\
                 K^- & \bar{K}^0 & -\frac{2\eta}{\sqrt{6}}\\
               \end{array}\right)
\end{equation}
as well as $\chi = 2B_0M_q\left(1 + \frac{A_0}{\sqrt{2}F}iP\right)$, 
$B_0\equiv\frac{m_{K}^2}{m_d+m_s}$ and $M_q$ is quark mass matrix. 
The last term in~\eqref{eq:a:4} describes the leading $\Delta I=\frac{1}{2}$
weak $\Delta S = 1$ transitions. Here $G_8 =
-\frac{G_F}{\sqrt{2}}V^*_{ud}V_{us}g_8$ with $g_8\simeq 
5.0$~\cite{Cirigliano:2011ny} and $\lambda = \frac{1}{2}(\lambda_6 +
i\lambda_7)$ where $\lambda_a$ are Gell-Mann matrices. Using this interaction
lagrangian one can calculate the amplitudes of $K^+\to\pi^+P$ and
$K_S\to\pi^0P$~\eqref{eq:2.8}, $K^\pm\to Pl\nu_l$~\eqref{eq:2.9} and
$P\to\gamma\gamma$~\eqref{eq:2.6} decays at LO of ChPT. We see that the
contributions from 
individual parameters $\kappa_u, \kappa_d$ and $\kappa_s$ cancel as they
should. We verified obtained expressions by comparing them in the limit
$A_0\to0$ with similar expressions obtained for the case of axion-like
particles after an appropriate redefinition of coupling
constants~\cite{Bauer:2020jbp,Bauer:2021wjo}.

\end{document}